\documentclass[twocolumn,showpacs,preprintnumbers,amsmath,amssymb]{revtex4}
\usepackage{amssymb}
\usepackage{amsfonts}

\usepackage{graphicx}
\usepackage{dcolumn}
\usepackage{bm}

\begin{document}

\title{A new quantum theory of gravity in the framework of general relativity}

\author{Chang-Yu Zhu$^1$}

\affiliation{%
$^1$Department of Physics, Zhengzhou University, Zhengzhou, Henan
450052, China
}%

\author{Heng Fan$^{2}$}
\affiliation{%
$^2$Institute of Physics, Chinese Academy of Sciences, Beijing
100190, China
}%
\date{\today}

\begin{abstract}
{\bf Observed physical phenomena can be described well by quantum
mechanics or general relativity \cite{einstein1,einstein2}. People
may try to find an unified fundamental theory  \cite{weinberg} which
mainly aims to merge gravity with quantum theory.
However, difficulty in merging those theories self-consistently
still exists, and no such theory is generally accepted. Here we try
to propose a quantum theory with space and time in symmetrical
positions in the framework of general relativity. In this theory,
Dirac matter fields, gauge fields and gravity field are formulated
in an unified way which satisf Dirac equation, Yang-Mills equation
and Einstein equation in operator form. This combines the quantum
mechanics and general relativity .}
\end{abstract}

\maketitle


\section{Introduction}
Two great discoveries of science in 20th century are the theory of
relativity \cite{einstein1,einstein2} and quantum mechanics. At
first it looked as if non-relativistic quantum mechanics would be
enough to explain the spectrum of the hydrogen atom. However, to
explain a finer structure in the spectrum due to the spin of the
electron, Dirac \cite{dirac} introduced his equation, the Dirac
equation,  by adding spin corrections to the Schr\"odinger equation
\cite{schrodinger}. Dirac equation satisfies invariance under
special relativity, which can describe a single-particle obeying
both relativity and quantum mechanics. This merges the theory of
special relativity to quantum mechanics.

Dirac equation is also a field theory describing ``field-like''
objects. Later, the quantum field theory provides a unified
framework describing both fields and particles
\cite{QFT1,QFT2,QFT3,QFT4}. In 1950s, Yang and Mills \cite{YM}
introduced a class of quantum field theories known as gauge
theories. In 1960s and 1970s, based on gauge theories, the Standard
Model
\cite{standard1,standard2,standard3,standard4,kmmatrix,standard5}
and related theories, see \cite{Nobel} for complete references, were
proposed for elementary particles and the interactions between them.
Up to then, the electromagnetic force, weak and strong nuclear
forces are merged into an unified model. While only gravity which is
one of the four basic forces in Nature remained outside of this
unified framework.

In the past decades, much effort has been put into studying how to
combine quantum mechanics with general relativity into a quantum
theory of gravity. In the Standard Model and field theory, the
motion of point-like particles is described by a graph of its
position against time which forms a world-line.

In this work, we start directly from the general relativity,
particles and fields are set of "events" which are quantum states of
quantum field. An event is like a world-point in 4D space-time, a
concept first coined by Einstein in which there is no absolute space
and time. We propose a quantum theory with general relativity, an
event or a quantum state is described by not only a complete set of
canonical coordinates or canonical momentums of 4D space in operator
form $\hat{x}^{\alpha }, \hat{p}_{\alpha }$, but also spin operators
$\hat {s}_{\mu \nu }$ and gauge charge operators $\hat {T}_{a}$.
These operators are covariance and constitute a Lie algebra. We then
define a covariance derivative operator. In contrast with the
derivative operator in Standard Model or in conventional quantum
field theory, this covariance derivative operator has frame and
connections of general relativity. Then the Einstein equation of
general relativity, Dirac equation and Yang-Mill equation are all
reformulated by this covariance derivative operator. This provides
an unified framework of quantum mechanics and general relativity.


\section{Elementary particles and quantum fields}
What we study is the all elementary particles and the interactions
between them. As usual, they are all named as fields in quantum
field theory. The elementary particles are divided into three
families as matter particles ($\Psi $), gauge particles ($A$) and
graviton ($g$). They satisfy Dirac equation, Yang-Mills equation and
Einstein equation, respectively. In this work, we will present an
unified quantum theory for those equations

Matter particles ($\Psi $) include quarks and leptons each class
have 6 types depending on mass and electro-charge. Each type of
lepton has isospin singlet state and isospin doublet state, so
leptons have 12 classes, altogether. Each flavor of quark has three
color states, each color has isospin singlet state and isospin
doublet state. So quarks have 36 types. Thus matter particles have
48 types. Note that the spin of matter particles is $1/2$ described
by a 4D Dirac spinor

Gauge particles ($A$) are divide into photons, three types of weak
gauge bosons and eight types of gluons. Photons have no mass, while
gluons is generally assumed to be massless, however, our theory
allows the massive gluons. The masses of weak gauge bosons are $m_Z,
m_{W_+}, m_{W_-}$ corresponding to three types of bosons $Z_0, W_+,
W_-$, respectively. We also have $m_{W_+}=m_{W_-}=m_Z\cos \theta
_W$, where $\theta _W$ is Weinberg angle. The spin of gauge
particles is 1 and thus are bosons. The mass of gluons are all
equal.

There is only one type of graviton ($g$), its mass and gauge charge
are zeroes, the spin is 2 in tensor representation.

There are 12 gauge charges, including hypercharge $\hat{Y}$, isospin
charges $\hat{I}_i, (i=1,2,3)$ and color charges $\hat{\lambda} _p,
(p=1,2,...,8)$. They satisfy the commutation relations
$[\hat{I}_i,\hat{I}_j]=i\epsilon _{ijk}\hat{I}_k$, $[\hat{\lambda
}_p,\hat{\lambda }_q]=if_{pq}^r\hat{\lambda }_r$, $[\hat {Y},
\hat{I}_i]=[\hat{Y}, \hat{\lambda}_p]=[\hat
{I}_i,\hat{\lambda}_p]=0$, where $\epsilon _{ijk}$ is Levi-Civita
symbol, $f_{pq}^r$ are structure constants of $su(3)$ group. We use
notations $\hat{t}_a, (a=1,2,...,12)$ to represent those 12 gauge
charges as $\hat{t}_1=g_1\frac {\hat {Y}}{2}$,
$\hat{t}_{1+i}=g_2\hat {I}_i, (i=1,2,3)$, $\hat {t}_{4+p}=g_3\hat
{\lambda }_p, (p=1,2,...,8)$, where $g_1,g_2$ and $g_3$ are
coefficients in Standard Model. So the commutation relations take a
concise form as $[\hat{t}_a,\hat{t}_b]=id_{ab}^c\hat {t_c}$, where
coefficients $d_{ab}^c$ are defined as
$d_{1+i,1+j}^{1+k}=g_2\epsilon _{ijk}$,
$d_{4+p,4+q}^{4+r}=g_2f_{pq}^r$ and $d_{ab}^c=0$ elsewhere. Those
gauge charges $\hat{t}_a$ can be denoted as gauge bosons in
Cartan-Weyl basis $\hat {T}_a$, (a=1,2,...,12) and satisfy the
relation $[\hat {T}_a, \hat {T}_b]=C_{ab}^c\hat {T}_c$.

In 4D space-time, we use notations $\hat {s}_{\alpha \beta }$,
$(\alpha ,\beta =0,1,2,3)$ to denote the spin operators which are
antisymmetric $\hat {s}_{\alpha \beta }=-\hat {s}_{\beta \alpha }$
and satisfy the commutation relation $[\hat {s}_{\alpha \beta },
\hat {s}_{\rho \sigma }]=-i(\eta _{\alpha \rho}\hat {s}_{\beta
\sigma }-\eta _{\beta \rho}\hat {s}_{\alpha \sigma }+\eta _{\alpha
\sigma}\hat {s}_{\rho \beta }-\eta _{\beta \sigma}\hat {s}_{\rho
\alpha })$, where $\eta _{\alpha \beta }$ is the Minkowski metric
defined as $\eta _{00}=1,\eta _{ij}=-\delta _{ij}, \eta _{0i}=\eta
_{i0}=0, \eta ^{\alpha \beta }=\eta _{\alpha \beta}$

Similar as in 3D quantum mechanics, in 4D space-time, the coordinate
and momentum are denoted as $\hat {x}^{\mu }$ and $\hat {p}^{\nu }$
which satisfy the relations $[\hat {x}^{\mu }, \hat {p}^{\nu
}]=-i\delta _{\nu }^{\mu }$, $[\hat {x}^{\mu }, \hat {x}^{\nu }]=
[\hat {p}_{\mu }, \hat {p}_{\nu }]=0$. Here we let the coordinates
and momentums be general covariance so that coordinates and
momentums transformation and inverse transformation take the forms
$\hat {x}'^{\mu }=\hat {x}'^{\mu }(\hat {x})$,$\hat {x}^{\mu }=\hat
{x}^{\mu }(\hat {x}')$ and $\hat {p}_{\mu }'=\frac {\partial \hat
{x}^{\nu }}{\partial \hat {x}'^{\mu }}\hat {p}_{\nu }$, $\hat
{p}_{\mu }=\frac {\partial \hat {x}'^{\nu }}{\partial \hat {x}^{\mu
}}\hat {p}'_{\nu }$. One can check that commutation relations remain
invariant under general coordinates transformation $[\hat {x}'^{\mu
}, \hat {p}'^{\nu }]=-i\delta _{\nu }^{\mu }$, $[\hat {x}'^{\mu },
\hat {x}'^{\nu }]= [\hat {p}'_{\mu }, \hat {p}'_{\nu }]=0$. In this
sense, we mean those commutation relations are quantum general
covariance. In our theory, all physical equations should be quantum
general covariance. This can be considered as the generalization of
the general covariance from classical case to quantum case and thus
can realize the unification between general relativity and quantum
mechanics. Under the condition of quantum general covariance, we
will present a quantum theory for gravity.

The coordinates and momentum are independent with spin and gauge
charges $[\hat {x}^{\mu }, \hat {s}_{\alpha \beta }]=[\hat {x}^{\mu
}, \hat {T}_a]=[\hat {p}_{\mu }, \hat {s}_{\alpha \beta }]= [\hat
{p}_{\mu }, \hat {T}_a]=[\hat {s}_{\alpha \beta }, \hat {T}_a]=0$.

\section{Representation theory and covariance derivative operator}
In our theory, field is described by spin, gauge charges and
coordinates-momentum. It is represented as a vector for matter
fields or operator for force fields in a space tensor product by
coordinate-momentum space, spin space and gauge space,
$V(M)=V_{xp}\otimes V_S(M)\otimes V_g(M)$. The quantum state
$|e_{st}\rangle $ of a matter particle is defined as the tensor
product in coordinate-basis, spin-basis and gauge-basis,
$|e_{st}(x)\rangle =|x\rangle \otimes |e_s\rangle \otimes
|e_t\rangle $, where $x\in R^4, s=1,2,3,4$ and $t=1,2,...,48$. It is
the common-eigenstate of 10 operators $\hat {x}^0,\hat {x}^1, \hat
{x}^2, \hat {x}^3, \hat {\gamma }_5,\hat {s}_{12}, \hat {Y},
\hat{I}_3, \hat {\lambda }_3$ and $\hat {\lambda }_8$. In this
representation, coordinate state can be changed to momentum state
$|x\rangle \rightarrow |p\rangle $ and we have $|e_{st}(p)\rangle
=|p\rangle \otimes |e_s\rangle \otimes |e_t\rangle $. The quantum
state of matter-particle $|\psi \rangle $ can then be expanded by
either coordinate state $|e_{st}(x)\rangle $ or momentum state
$|e_{st}(p)\rangle $,
\begin{eqnarray}
|\Psi \rangle =\int _{R^4}\Psi ^{st}(x)|e_{st}(x)\rangle d^4x =\int
_{R^4}\tilde {\Psi }^{st}(p)|e_{st}(p)\rangle d^4p,
\end{eqnarray}
where coefficients $\Psi ^{st}(x)$ in the expansion are defined as
$\Psi ^{st}(x)=\langle e^{st}(x)|\Psi \rangle =(2\pi )^{-2}\int
_{R^4}\tilde {\Psi }^{st}(p)\exp (-ipx)d^4p$, similar for case
$\tilde {\Psi } ^{st}(p)$. Please note in our representation, spin,
gauge and general coordinate-momentum are dealt in the symmetric
positions. In quantum mechanics, the time evolution of a quantum
state or an operator are described by Sch\"odinger representation or
Heisenberg representation, respectively. We may notice that time and
space are not symmetric. In comparison for our work, there is no
absolute time and space in general relativity, thus four coordinates
are dealt symmetrically. The state $|e_{st}\rangle $ is an event.

The spin frame formalism takes the form $\hat{\theta }=\hat{\theta
}_{\alpha }^{\mu }\otimes \hat {\gamma }^{\alpha }\otimes \hat
{p}_{\mu }$, $\hat{\theta }_{\alpha }^{\mu }=\int _{R^4}\tilde
{\theta }_{\alpha }^{\mu }(x)\hat {\varepsilon }(x)d^4x=\int
_{R^4}\tilde {\theta }_{\alpha }^{\mu }(p)\hat {\varepsilon
}(p)d^4p$, where $\hat {\gamma }^{\alpha }$ are Dirac matrices,
$\hat {\varepsilon }(x)=\delta ^4(\hat {x}-x)=\int_{R^4}\delta
^4(x'-x)|x'\rangle \langle x'|d^4x'=|x\rangle \langle x|$, $\hat
{\theta }_{\alpha }^{\mu }$ is the coefficient, $\theta _{\alpha
}^{\mu }(x)$ and $\tilde {\theta }_{\alpha }^{\mu }(x)$ are
coordinate functions and momentum functions in spin frame formalism,
they satisfy relations $\tilde {\theta }_{\alpha }^{\mu }(p)=(2\pi
)^{-2}\int _{R^4}\theta _{\alpha }^{\mu }(x)\exp (ipx)d^4x$, $\theta
_{\alpha }^{\mu }(p)=(2\pi )^{-2}\int _{R^4}\tilde {\theta }_{\alpha
}^{\mu }(p)\exp (-ipx)d^4p$. The spin frame formalism $\hat {\theta
} _{\alpha }=\hat {\theta }_{\alpha }^{\mu }\otimes \hat {p}_{\mu }$
satisfy the commutation relations $[\hat {\theta }_{\alpha }, \hat
{\theta }_{\beta }]=i\hat {f}_{\alpha \beta }^{\gamma }\hat {\theta
}_{\gamma }$, where $\hat {f}_{\alpha \beta }^{\gamma }$ are the
structure coefficients in spin frame formalism and is represented as
$\hat {f}_{\alpha \beta }^{\gamma }=(\hat {\theta }_{\alpha }^{\mu
}\partial _{\mu }\hat {\theta }_{\beta }^{\nu }-\hat {\theta
}_{\beta }^{\mu }\partial _{\mu }\hat {\theta }_{\alpha }^{\nu
})\hat {\theta }_{\nu }^{\gamma }$. The gravity connection is
defined as $\hat {\Gamma }=\frac {1}{2}\hat {\Gamma }_{\alpha
}^{\rho \sigma }\otimes \hat {\gamma }^{\alpha }\otimes \hat
{s}_{\rho \sigma }$, where $\hat {\Gamma }_{\alpha }^{\rho \sigma
}=\frac {1}{2}(\hat {f}_{\alpha }^{\rho \sigma }+\hat {f}_{\alpha
}^{\rho ,\sigma }-\hat {f}_{\alpha }^{\sigma ,\rho })$, and we have
used the notations $\hat {f}_{\tau }^{\rho \sigma }=\eta ^{\rho
\alpha }\eta ^{\sigma \beta }\eta _{\tau \gamma }\hat {f}_{\alpha
\beta }^{\gamma }$, $\hat {f}_{\tau }^{\rho ,\sigma }=\eta ^{\sigma
\beta }\hat {f}_{\tau \beta }^{\rho }$.

The gauge connections is defined as $\hat {A}=\hat {A}_{\alpha
}^a\otimes \hat {\gamma }^{\alpha }\otimes \hat {T}_a$, similar as
for gravity field we have $\hat {A}_{\alpha }^a=\int _{R^4}A_{\alpha
}^a(x)\hat {\varepsilon }(x)d^4x=\int _{R^4}\tilde {A}_{\alpha
}^a(p)\hat {\varepsilon }(p)d^4p$, where $A_{\alpha }^a(x)$ and
$A_{\alpha }^a(x)$ are coordinate and momentum functions,
respectively.

Now we define the covariance derivative operator as
\begin{eqnarray}
\hat {D}_{\alpha }=-i\hat {\theta }_{\alpha }^{\mu }\otimes \hat
{p}_{\mu }+\frac {i}{2}\hat {\Gamma }_{\alpha }^{\rho \sigma
}\otimes \hat {s}_{\rho \sigma }-i\hat {A}_{\alpha }^a\otimes \hat
{T}_a
\end{eqnarray}
This operator has connections of gravity, connections of gauge and
spin frame. Thus it can describe all force fields. As in quantum
mechanics, when acting on operators,  it is represented in the form
of commutating calculation, when acting on matter fields, it is
represented as an operator acting on quantum states. The covariance
differential can take the form $\hat {D}=\hat {\gamma }^{\alpha
}\otimes \hat {D}_{\alpha }$.

We then define the interaction curvature as
\begin{eqnarray}
\hat {\Omega }_{\alpha \beta }=i[\hat {D}_{\alpha }, \hat {D}_{\beta
}]-i\hat {f}_{\alpha \beta }^{\gamma }\hat {D}_{\gamma },
\end{eqnarray}
and $\hat {\Omega }=\hat {\Omega }_{\alpha \beta }\otimes \hat
{s}^{\alpha \beta }$. It is the summation of gravity curvature and
gauge curvature $\hat {\Omega }_{\alpha \beta }=\frac {1}{2}\hat
{R}_{\alpha \beta }^{\rho \sigma }\otimes \hat {s}_{\rho \sigma
}+\hat {F}_{\alpha \beta }^a\otimes \hat {T}_a$. The gravity
curvature takes the form $\hat {R}_{\alpha \beta }^{\rho \sigma
}=\hat {e}_{\alpha }^{\mu }\partial _{\mu }\hat {\Gamma }_{\beta
}^{\rho \sigma }-\hat {e}_{\beta }^{\mu }\partial _{\mu }\hat
{\Gamma }_{\alpha }^{\rho \sigma }+\hat {\Gamma }_{\gamma ,\alpha
}^{\rho }\hat {\Gamma }_{\beta }^{\gamma \sigma }-\hat {\Gamma
}_{\gamma ,\beta }^{\rho }\hat {\Gamma }_{\alpha }^{\gamma \sigma
}-\hat {f}_{\alpha \beta }^{\gamma }\hat {\Gamma }_{\gamma }^{\rho
\sigma }$, and the gauge curvature takes the form $\hat {F}_{\alpha
\beta }^{a}=\hat {e}_{\alpha }^{\mu }\partial _{\mu }\hat {A}_{\beta
}^a-\hat {e}_{\beta }^{\mu }\partial _{\mu }\hat {A}_{\alpha
}^a+C_{bc}^a\hat {A}_{\alpha }^b\hat {A}_{\beta }^c -\hat
{f}_{\alpha \beta }^{\gamma }\hat {A}_{\gamma }^a$.

The covariance derivative operator and the interaction curvature
satisfy the Bianchi identity, $\hat {D}_{\alpha }\hat {\Omega
}_{\beta \gamma }+\hat {D}_{\beta }\hat {\Omega }_{\gamma \alpha
}+\hat {D}_{\gamma }\hat {\Omega }_{\alpha \beta }=0$. It turns out
be Bianchi identities for gravity curvature and  gauge curvature,
respectively, $\hat {D}_{\alpha }\hat {R}_{\beta \gamma }^{\rho
\sigma }+\hat {D}_{\beta }\hat {R }_{\gamma \alpha }^{\rho \sigma
}+\hat {D}_{\gamma }\hat {R}_{\alpha \beta }^{\rho \sigma }=0$ and
$\hat {D}_{\alpha }\hat {F}_{\beta \gamma }^{a}+\hat {D}_{\beta
}\hat {F}_{\gamma \alpha }^{a}+\hat {D}_{\gamma }\hat {F}_{\alpha
\beta }^a=0$.

\section{Dirac equation, Yang-Mills equation and Einstein equation}
The Dirac equation for matter fields can be written as:
\begin{eqnarray}
(i\hat {\gamma }^{\alpha }\hat {D}_{\alpha }-\hat {m})|\Psi \rangle
=0,
\end{eqnarray}
where $\hat {m}$ is the mass matrix in gauge space, its eigenvalues
are masses of the corresponding elementary particles.

The Yang-Mills equation takes the form
\begin{eqnarray}
\hat {D}_{\alpha }\hat {F}_a^{\alpha \beta }=\hat {j}_a^{\beta
}+M_a^b\hat {A}_b^{\beta },
\end{eqnarray}
where the l.h.s is $\hat {D}_{\alpha }\hat {F}_a^{\alpha \beta
}=i\hat {\theta }_{\alpha }^{\mu }[\hat {p}_{\mu }, \hat
{F}_a^{\alpha \beta }]+\hat {\Gamma }^{\alpha }_{\rho ,\alpha }\hat
{F}_{a}^{\rho \beta }-\hat {\Gamma }^{\beta }_{\rho ,\alpha }\hat
{F}_{a}^{\rho \alpha }+C_{ab}^c\hat {A}^b_{\alpha }\hat
{F}_c^{\alpha \beta }$, the r.h.s is the total current density,
$\hat {j}_a^{\beta }$ is gauge current density , $M_a^b$ are mass
tensor of gauge bosons with $M_2^2=m_z^2$, $M_3^3=M_4^4=m^2_{W_{\pm
}}$, and $M^{4+p}_{4+p}=m, p=1,2,\cdots ,8$, $M_a^b=0$ elsewhere,
the mass of gluon is $m$.

The Einstein equation takes the form,
\begin{eqnarray}
\hat {R}^{\alpha }_{\beta }-\frac {1}{2}\delta _{\beta }^{\alpha
}\hat{R}=-8\pi G\hat {T}^{\alpha }_{\beta },
\end{eqnarray}
where $G$ is gravity constant, $\hat {T}^{\alpha }_{\beta }$ is
energy-momentum tensor, $\hat {R}^{\alpha }_{\beta }=\hat
{R}^{\alpha \gamma }_{\beta \gamma }$ and $\hat {R}=\hat {R}^{\alpha
}_{\beta }$. The Einstein equation and the contracted tensor of
Bianchi identity of gravity $\hat {D}_{\alpha }(\hat {R}^{\alpha
}_{\beta }-\frac {1}{2}\delta ^{\alpha }_{\beta }\hat {R})=0$ can
lead to the energy-momentum conservation law, $\hat {D}_{\alpha
}\hat {T}^{\alpha }_{\beta }=0$.

\section{Physical quantities and energy-momentum conservation law}
Current density of particles is defined as $\rho ^{\alpha
}(x)=\langle \Psi |\hat {\varepsilon}(x)\hat {\gamma }^{\alpha
}|\Psi \rangle $. Due to Dirac equation, the conservation of the
number of particles can be found to be $\hat {D}_{\alpha }\rho
^{\alpha }=0$. For each $t$ in gauge basis, the current density of
the particle takes the form $\rho ^{\alpha }_t(x)=\langle \Psi _t
|\hat {\varepsilon}(x)\hat {\gamma }^{\alpha }|\Psi _t\rangle $.

Gauge current density of the matter fields is defined as $\hat
{j}_a^{\alpha }(x)=\langle \Psi |\hat {\varepsilon}(x)\hat {\gamma
}^{\alpha }\hat {T}_a|\Psi \rangle $, gauge charge density of
production rate of matter fields takes the form $\hat
{u}_a(x)=-i\langle \Psi |[\hat {T}_a, \hat {m}]\hat
{\varepsilon}(x)|\Psi \rangle $, by Dirac equation we have $\hat
{D}_{\alpha }\hat {j}_a^{\alpha }=\hat {u}_a$. For electric-charge
and color charges $\hat {u}_a=0$, they are conserved quantities. For
three type weak charges $\hat {u}_a\not= 0$.

The spin angular-momentum current density of the matter fields take
the form $S^{\alpha \beta \gamma }(x)=\frac {1}{2}\langle \Psi |\hat
{\varepsilon}(x)(\hat {\gamma }^{\alpha }\hat {s}^{\beta \gamma
}+\hat {s}^{\beta \gamma }\hat {\gamma }^{\alpha })|\Psi \rangle $,
they are anti-symmetric. And one may check $\hat {D}_{\alpha
}S^{\alpha \beta \gamma }=0$.

The energy-momentum tensor of the matter fields is defined as $\hat
{t}^{\alpha }_{\beta }(x)=\frac {i}{4}\{\langle |\Psi |\hat
{\varepsilon}(x)[\hat {\gamma }^{\alpha }\hat {D}_{\beta }+\hat
{\gamma }_{\beta }\hat {D}^{\alpha })|\Psi \rangle ]+[\langle \Psi
|(\hat {D}_{\beta }\hat {\gamma }^{\alpha }+\hat {D}^{\alpha }\hat
{\gamma }_{\beta })]\hat {\varepsilon}(x)|\Psi \rangle \}$.
According to Dirac equation, the divergence equation for
energy-momentum tensor can be proved to be $\hat {D}_{\alpha }\hat
{t}^{\alpha }_{\beta }=\hat {F}^a_{\alpha \beta }\hat {j}^{\alpha
}_a+A^a_{\beta }\hat {u}_a$, where we can find gauge current density
and the density of production rate appear in the r.h.s, while the
gravity field is not involved.

The energy-momentum tensor of gauge fields is defined as $\hat {\tau
}^{\alpha }_{\beta }=\hat {F}_a^{\alpha \rho }\hat {F}^a_{\beta \rho
}-\frac {1}{4}\delta ^{\alpha }_{\beta }\hat {F}_a^{\rho \sigma
}\hat {F}^a_{\rho \sigma }+M^a_b\hat {A}_a^{\alpha }\hat
{A}^b_{\beta }-\frac {1}{2}\delta ^{\alpha }_{\beta }M^a_b\hat
{A}_a^{\rho }\hat {A}^b_{\rho }$. By Yang-Mills equation and Bianchi
identity, we find the divergence of the energy-momentum of gauge
fields be $\hat {D}_{\mu }\hat {\tau }^{\alpha }_{\beta }=-\hat
{j}_a^{\alpha }\hat {F}^a_{\alpha \beta }-\hat {A}^a_{\beta }\hat
{u}_a$.

The total energy-momentum tensor is the summation of the
energy-momentum tensors of matter fields and gauge fields $\hat
{T}^{\alpha }_{\beta }=\hat {t}^{\alpha }_{\beta }+\hat {\tau
}^{\alpha }_{\beta }$. We thus find that the energy-momentum
conservation law, $\hat {D}_{\alpha }\hat {T}^{\alpha }_{\beta }=0$,
can be also proved by Yang-Mills equation and Dirac equation. So
energy-momentum conservation law proved by two different ways shows
that the compatible of our theory in presenting the three
fundamental equations, Einstein equation, Dirac equation and
Yang-Mills equation.

\section{Outlook}
A combined and unified description of quantum mechanics and general
relativity is very important for physics. It is not only for a
fundamental understanding of nature but also it can provide a theory
for systems where both quantum mechanics and general relativity is
important. We will see in Ref.\cite{ZF0} that with this unified
theory we can provide a solution of dark energy of cosmos which is
one of the most mysterious problem presently since the dark energy
is the main part (around $72\%$) of our universe. While the dark
energy solution will, on the other hand, predict that the elementary
particle gluon are massive though its mass is very small. We will
also see that this theory will lead to some exciting results such
as: mass problem is solved for gauge theory, we can explain the
color confinement of quarks, the parity violation for weak
interactions, we can find that gravity can cause the CPT violation.
Conceptually, we will find that no Higgs mechanism is necessary,
while all of our results agree with the basic facts of physics. The
detailed presentation of the whole theory is in Ref.\cite{ZF0}.

We believe that this theory can provide a foundation of quantum
physics. Thus a lot of problems should be clarified, in particular,
our results should agree with all well-established results both
theoretically and experimentally.

\acknowledgements

We thank Mr. Shi-Ping Ding for consistent supporting and
discussions.  This work is supported by grants of National Natural
Science Foundation of China (NSFC) Nos.(10674162,10974247), ``973''
program (2010CB922904) of Ministry of Science and Technology (MOST),
China, and Hundred-Talent Project of Chinese Academy of Sciences
(CAS), China.

\end{document}